\documentclass{article}

\usepackage[preprint, nonatbib]{neurips_2023}

\usepackage[utf8]{inputenc} 
\usepackage[T1]{fontenc}    
\usepackage{hyperref}       
\usepackage{url}            
\usepackage{booktabs}       
\usepackage{amsfonts}       
\usepackage{nicefrac}       
\usepackage{microtype}      
\usepackage{xcolor}         
\usepackage{biblatex}       
\usepackage{amsmath}
\usepackage{bbm}
\usepackage{graphicx}
\usepackage{caption}
\usepackage{subcaption}

\newcommand{\norm}[1] { \left\Vert #1 \right\Vert }

\title{\textit{Noisy Neighbors}: Efficient membership inference attacks against LLMs}

\author{%
  Filippo Galli\thanks{Part of this author's work was carried out while at Meta Inc.} \\
  Scuola Normale Superiore\\
  Scuola Superiore Sant'Anna\\
  Pisa, Italy\\
  \And
  Luca Melis \\
  Meta Inc. \\
  \And
  Tommaso Cucinotta \\
  Scuola Superiore Sant'Anna\\
  Pisa, Italy \\
}

\begin{document}

\maketitle

\begin{abstract} 
The potential of transformer-based LLMs risks being hindered by privacy concerns due to 
their reliance on extensive datasets, possibly including sensitive information. 
Regulatory measures like GDPR and CCPA call for using robust auditing tools to address 
potential privacy issues, with Membership Inference Attacks (MIA) being the primary 
method for assessing LLMs' privacy risks. Differently from traditional MIA approaches, 
often requiring computationally intensive training of additional models, this paper introduces an efficient methodology that generates \textit{noisy neighbors} for a target sample by adding stochastic noise in the embedding space, requiring operating the target model in inference mode only. Our findings demonstrate that this approach closely matches the effectiveness of employing shadow models, showing its usability in practical privacy auditing scenarios.
\end{abstract}

\section{Introduction}

Advancements in natural language processing \cite{vaswani2017attention} have made large 
language models (LLMs) \cite{radford2019language} essential for many text tasks. However, 
LLMs face issues like biases \cite{narayanan2023nationality}, privacy breaches 
\cite{carlini2021extracting}, and vulnerabilities \cite{wallace2021concealed}, underscoring 
the importance of protecting user privacy. The use of large datasets including personal 
information, has raised privacy concerns, leading to regulations such as GDPR \cite{gdpr} 
and CCPA \cite{ccpa}.

Membership inference attacks (MIA) \cite{shokri2017membership} are effectiv
auditing tools aiming at determining if a specific data point was used in an LLM's 
training dataset by analyzing its output. Such attacks highlight potential privacy 
breaches, relying on models' tendency to overfit to familiar data 
\cite{carlini2019secret}. By employing calibration strategies and training shadow 
models, the accuracy of MIAs can be improved, although challenges such as 
computational demands and limitations in effectiveness when deviating from training 
distribution assumptions persist. In this paper, we contribute to this field by: 
i) exploring membership inference attacks from the 
standpoint of a privacy auditor, ii) introducing a 
computationally efficient calibration strategy that sidesteps training shadow models,
and iii) empirically assessing its potential in replacing
other prevalent strategies.

\section{Background}

LLMs generate a probability distribution over their vocabulary based on a tokenized input 
sequence converted into numerical inputs through an embedding layer. This layer maps tokens 
to a dense representation, which can be learned during training \cite{radford2018improving, 
radford2019language} or derived from public \textit{word embeddings} \cite{devlin2018bert}.
For a model $f$ with input sequence $x$, we define 
$\mathbb{P}[w| x] = f_w(x)$ as the conditional 
probability that the token following $x$
is $w$.
LLMs are typically trained on large datasets of text to minimize
a measure of surprise in seeing the next token, called 
\textit{perplexity}. For a sequence $x$, it is defined as the average negative
log-likelihood of its tokens:
\begin{equation}
  ppx(f, x) = -\frac{1}{|x|} \sum_{t = 1}^{|x|}\log (f_{x_t}(x_{<t}))
\end{equation}
with $|x|$ the number of tokens in the sequence.

Membership inference attacks \cite{shokri2017membership, watson2021importance,
carlini_membership_2022} 
aim to determine whether a particular data record $x$ was
used in the training dataset $D_{train}$ of a machine learning model.
These methods leverage model outputs like confidence scores or prediction
probabilities to compute a score for the targeted sample. 
For LLMs, the typical assumption is to grant the adversary access to
the output probabilities $f(x)$, which may be used to estimate the perplexity on
the targeted samples as a score. 
Given a sample $x$, the goal of the attacker is to learn a thresholding classifier
to output $1$ when the perplexity is lower than a certain value $\gamma$:
\begin{equation}
  A_{\gamma}(f, x) = \mathbbm{1}[ppx(f, x) < \gamma]
\end{equation}
MIA is a simple and effective tool to measure the privacy risk in a trained machine 
learning model, and it has interesting connections with other privacy frameworks. In
particular, it is known to have a success rate bounded by the privacy parameters of 
Differential Privacy (DP) \cite{dwork2006calibrating}. A randomized mechanism 
$\mathcal{M}$ is said to be $\varepsilon$-DP if for any two datasets 
$D, D'$ that differ in at most one sample, and for any
$R \subseteq \text{range}(\mathcal{M})$, we have:
\begin{equation}
  \mathbb{P}[\mathcal{M}(D) \in R] \leq e^{\varepsilon} 
  \mathbb{P}[\mathcal{M}(D') \in R]
\end{equation}
Notably, DP quantifies the worst-case scenario of the privacy risk, 
so it is a fundamental tool in privacy assessment. From the performance of the
thresholding classifier $\tilde{A}_{\gamma}(f, x)$ one can obtain a lower bound to
the \textit{empirical} $\varepsilon$-DP \cite{kairouz_composition_2015}:
\begin{equation} \label{edp}
  e^{\varepsilon} \geq \frac{TPR}{FPR}
\end{equation}
with TPR and FPR being, respectively, the true and false positive rates, given a 
certain threshold.

\section{Related works}
Privacy attacks against language models
is an active area of research and different refinements have been proposed.
Some works have focused on an attacker where data poisoning is allowed, granting 
the adversary write access to the training dataset, to increase memorization 
\cite{tramer2022truth} or in general to induce malicious behaviours 
\cite{xu2023instructions, wallace2021concealed, yan2023virtual,shu2024exploitability,
huang2020metapoison} and improve property inference attacks \cite{mahloujifar2022property}.
Other works have adopted similar techniques to achieve actual training data extraction from
the training set, with only query access to the trained model 
\cite{carlini2021extracting, carlini2023extracting}. 

In the context of MIAs with query 
access to the target model, most research focused on strategies to improve the calibration
of the per-sample scores, i.e. techniques to improve the precision and recall in 
distinguishing members from non-members of the training set.
In principle, if we can assert that an out-of-distribution
non-member of the training set will induce a high perplexity in a target LLM,
there are a number of scenarios where
the distinction is not as clear cut, and a thresholding classifier essentially ends up 
distinguishing between in-distribution from out-of-distribution samples. A refined MIA
then employs calibration strategies to tune the scoring function based on the
difficulty of classifying the specific sample, as in \cite{watson2021importance}. 
Thus, a relative membership score is obtained by comparing $f(x)$ with one of two
results based on whether the
adversary is assumed to have access to \textit{neighboring models} 
$\tilde{f}(x)$ \cite{carlini_membership_2022, watson2021importance} 
or \textit{neighboring samples} $f(\tilde{x})$ \cite{mattern2023membership}. 
The new classifier becomes:
\begin{equation}
  \tilde{A}_{\gamma}(f, x) = \mathbbm{1}[ppx(f, x) - \tilde{ppx}(f, x) < \gamma]
\end{equation}
where $\tilde{ppx}(f, x)$ is the calibrated score over a set of neighboring models 
$ppx(\tilde{f}, x)$ or over a set of neighboring samples $ppx(f, \tilde{x})$.
Neighboring models can be obtained by an adversary who is assumed to have some degree of
knowledge of the training data distribution and trains a number of shadow models to
mimic the behaviour of the target LLM. For instance \cite{carlini_membership_2022}
trains multiple instances of the same
architecture on different partitions of the training set, 
\cite{carlini2021extracting} uses smaller 
architectures trained on roughly the same data,
\cite{watson2021importance} leverages catastrophic forgetting 
of the target model under the assumption of white-box access.
Neighboring samples do not require this assumption nor additional training 
and only need a strategy to craft inputs that are similar to the target
sample under a certain distance metric. For instance, \cite{mattern2023membership}
crafts neighboring sentences by swapping a number of words with their synonyms, 
showing good results but applicable primarily when the adversary has limited knowledge of the 
training data distribution. The authors then base the neighboring relationship in the
\textit{semantic} space, which is hard to quantify and fix, resulting in the need to 
generate a large number of neighbors to reduce the effects of these random fluctuations.
Additionally, we emphasize how \cite{mattern2023membership} 
requires the use of an additional BERT-like model to generate synonyms, 
thus increasing the computational and memory cost of the attack.
In \cite{tramer2022truth} instead, calibration is done by comparing scores of the
true inputs with scores of the lower-cased inputs. These strategies are known to be 
under-performing when knowledge of the training distribution is available, and are
therefore proposed as an effective calibration mechanism when training shadow models
is not possible.

\section{Method}
\begin{figure}[ht]
    \centering
    \includegraphics[scale=0.60]{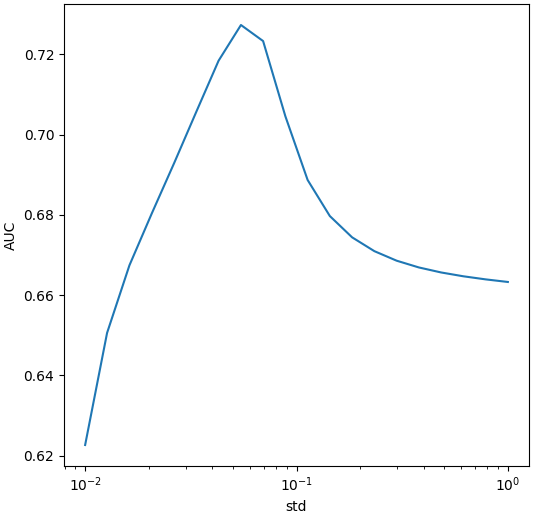}
    \caption{The AUC of the thresholding classifier for MIA shows a 
    single and prominent peak at the optimal $\sigma$ value in the 
    \textit{noisy neighbors} strategy.}
    \label{fig:peak}
\end{figure}

The intuition behind noisy neighbors is that, fixed a distance
from a sample, the target model will show a larger difference in perplexity
between a training sample and its neighbors than between a test sample and
its neighbors. Thus, if we describe a language model as a composition of layers
$f(x) = g(e(x))$ where $e$ is an embedding layer and $g$ is the rest of the network,
one can artificially create neighbors in
the $n$-dimensional embedding space by directly injecting random noise at
the output of $e(x)$. In particular, if we create noisy neighbors by injecting
Gaussian noise such that 
\begin{equation}
    f(x_{\sigma}') = g(e(x) + \rho),\quad \text{with}\; \rho \sim \mathcal{N}(0, \sigma I_n)
\end{equation}
then the Euclidean distance between the true and randomized input in the 
embedding space will be
\begin{equation}
   \mathbb{E}[\norm{e(x) - e(x) - \rho}] = \mathbb{E}[\norm{\rho}] = \sigma \sqrt{n}
\end{equation}
thus fixing, in expectation, the distance from the true sample at which the perplexity of 
the models will be evaluated. 
Generating multiple neighbors for each sample is crucial to mitigate randomness from 
stochastic noise, requiring repeated LLM inferences. Choosing the standard deviation 
$\sigma$ potentially involves a complex parameter search with many model queries. 
However, the strategy's performance shows a clear peak at the optimal $\sigma$ value,
as shown in Figure \ref{fig:peak}, which can be efficiently identified using binary search.

We emphasize the challenge of isolating the embedding layer from the remainder of the
network in an LLM when considering a scenario where an attacker has only 
black box access to the model. However, when this limitation does not apply, we think it is
still within the capacity of an auditor to utilize a slightly stronger attacker model,
where the first embedding layer is exposed, to save computational resources
in simulating an adversary without access to the model architecture.
Most importantly, in fact, we are inclined to explore this option as a more 
computationally efficient substitute for training shadow models for calibration, 
particularly in the context of auditing, rather than viewing it as a novel, 
realistic attack.

\section{Experiments}
\begin{figure}[ht]
  \begin{subfigure}[b]{0.48\textwidth}
  \centering
    \includegraphics[scale=0.52]{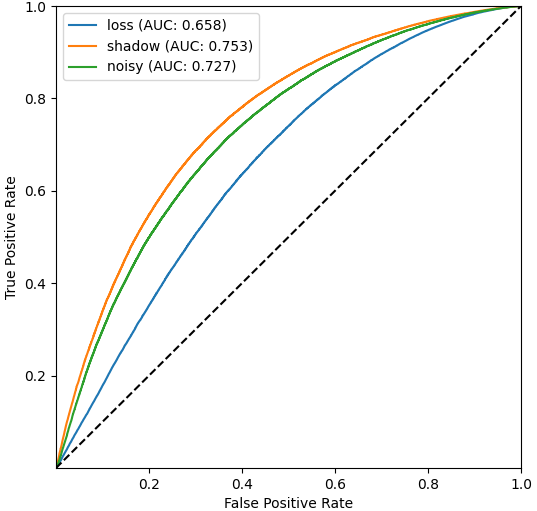}
    \caption{ROC curve of the MIA classifier.}
    \label{fig:roc}
  \end{subfigure}
  \hfill
  \begin{subfigure}[b]{0.48\textwidth}
  \centering
    \includegraphics[scale=0.52]{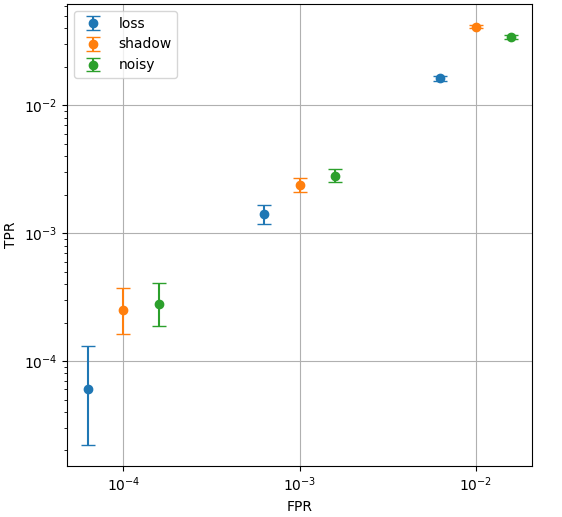}
    \caption{Performance of the attacker at low FPRs.}
    \label{fig:low_fpr}
  \end{subfigure}
  \caption{Efficacy of different strategies for MIA. Confidence intervals computed
  with the Clopper-Person exact method.}
  \label{fig:classifier}
\end{figure}

\begin{figure}
    \centering
    \includegraphics[scale=0.52]{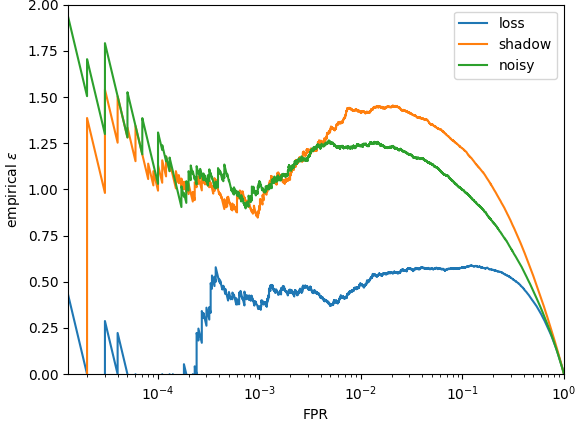}
    \caption{Empirical differential privacy measured downstream of training.}
    \label{fig:edp}
\end{figure}

To validate the noisy neighbor strategy in implementing a calibrated
MIA, we run a series of preliminary experiments on an LLM to gauge the risk of 
memorization of training data. The chosen architecture is GPT-2 \textit{small}
\cite{radford2019language}
to compromise learning capacity with memory and computational footprint at about $1.5$ 
billion parameters, especially
considering that competing strategies require training multiple LLMs from scratch.
The model was pre-trained on OpenWebText
\cite{Gokaslan2019OpenWeb}, an open reproduction of the undisclosed WebText in 
\cite{radford2019language}. The model was then fine-tuned on $60\%$ of the full WikiText 
corpus \cite{merity2016pointer}, a large collection of Wikipedia articles.
The same data split was then partitioned in $10$ subsets used to train $10$
shadow models for score calibration, as in \cite{carlini_membership_2022}.
Note that Wikipedia articles are filtered out of the
OpenWebtext corpus, to avoid data leakage in common benchmarks, such as ours. 
The remaining portion of $40\%$ of WikiText is thus used as source of 
non-member, $126$-token long samples to analyze the performance of the attack. 
We generate only $10$ synthetic neighbors for each sample.
Given a sample and its score, the thresholding classifier yields a binary decision on 
whether it was part of the training dataset or not. To determine how good the 
best possible classifier may be, we need to evaluate its accuracy at different 
thresholds. As it is common for binary classification problems, though, the accuracy
does not give a complete picture of the confidence at which the classifier is able to
tell apart members and non-members of the dataset. Thus Figure \ref{fig:roc} shows
the complete range of TPRs versus FPRs for the three main strategies we included in this 
comparison: score by perplexity (\textit{loss}),
by shadow model calibration (\textit{shadow}), and by noisy neighbor (\textit{noisy}) 
calibration. 
We have opted not to incorporate the \textit{lowercasing} 
strategy \cite{tramer2022truth} and the \textit{semantic 
neighbor} approach \cite{mattern2023membership} in our study. 
These methods have, however, shown lower performance levels
when information about the training data distribution is 
accessible, which is contemplated from the auditor point of 
view. Additionally, we faced challenges 
replicating some results from \cite{mattern2023membership}, 
possibly due to limitations in the synonym generation technique described
in \cite{zhou-etal-2019-bert}.
Figure \ref{fig:roc} also notes the Area Under the Curve (AUC), which 
for \textit{noisy} and \textit{shadow} amounts to $0.727$ and $0.753$ respectively,
thus showing a discrepancy of only $\sim 3.4\%$. The AUC is an important metric for
binary classifiers as it abstracts from the specific threshold, thus giving an average-case
idea of the strength of the attacker. Still, as highlighted in 
\cite{carlini_membership_2022}, special care should be given to what happens at low 
FPRs, that is when the attacker can confidently recognize members of the training set.
This is what Figure \ref{fig:low_fpr} focuses on, again showing a strong overlap of the
\textit{shadow} and \textit{noisy} strategies. Following Equation \ref{edp}, we also 
provide the perspective of empirical DP, as the privacy community pushes 
to adopt this framework to comply to regulatory frameworks such as the GDPR 
\cite{cummings2018role}. Empirical DP measures the extent to which individual data
points can be inferred or re-identified from the output of the system, and contrary to
DP, it is a \textit{post-hoc} measurement, not an \textit{a-priori} guarantee.
Figure \ref{fig:edp} reports the results, where we see a
strong consistency between the \textit{noisy} and \textit{shadow} strategies, especially 
for FPRs lower than $10^{-2}$.


\section{Limitations}
The effectiveness of the noisy neighbors method depends on assumptions that may 
not apply universally across models or datasets. Its success also relies on 
specific noise parameters, potentially limiting its generalizability. Despite 
being computationally more efficient than shadow model methods, it still requires 
significant computational resources.

\section{Conclusion}
This work set out to elaborate a strategy for membership inference attacks. Differently
from prior research 
focusing on improving the strength of the attacker, we develop
a technique trying to achieve a similar efficacy, while 
reducing
the computational burden for an auditor trying to assess the privacy risk of exposing
the query access to a trained LLM.
We propose the use of noise injection in the embedding space of the LLM
to create synthetic neighbors of the targeted sample, to shift the comparison from the 
perplexity scored by different models on one sample,
to the comparison of different samples by the same model.
This approach allows to only use the model in inference mode, thus inherently 
reducing the time and cost of running an MIA. With a number of experiments we assess 
how our strategy results converge to the results of using shadow models, showing a
remarkable alignment.

\printbibliography


\end{document}